\documentclass[11pt,a4paper]{article}

\pdfoutput=1   
\usepackage{amsmath,amsfonts,amssymb,amsbsy}
\usepackage{mathrsfs,latexsym}
\usepackage{graphicx}
\usepackage{color}
\usepackage{booktabs}
\usepackage{multirow}
\usepackage{multicol}
\usepackage{subfig}
\usepackage{alpha}
\hypersetup{%
  pdftitle = {How much do charm sea quarks affect the charmonium spectrum?},
  pdfauthor = { },
  pdfkeywords = { }
}%
\usepackage{macros_alpha}
\usepackage{textcomp}

\newcommand{\Op}{\mathcal{O}} 
\newcommand{\mcr}{m_{\rm cr}}
\newcommand{\LambdaMS}{\Lambda_{\overline{\rm MS}}}
\newcommand{\mPCAC}{m_{\rm PCAC}}

\newcommand{\Prop}{{\mathcal S}}
\newcommand{\tmPstar}{1.807463} 

\begin{document}

\preprintno{%
WUB/19-03
}

\title{How much do charm sea quarks affect the charmonium spectrum?}

\collaboration{\includegraphics[width=2.8cm]{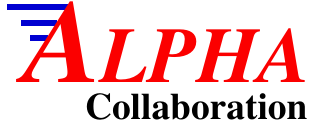}}

\author[ucy,wup]{ Salvatore Cal\`i}
\author[wup]{Francesco~Knechtli}
\author[wup]{Tomasz Korzec}

\address[ucy]{Department of Physics, University of Cyprus, 
              P.O. Box 20537,1678 Nicosia, Cyprus}
\address[wup]{Department of Physics, Bergische Universit{\"a}t Wuppertal, 
              Gaussstr.~20, 42119~Wuppertal, Germany}

\begin{abstract}
The properties of charmonium states are or will be intensively studied by the
B-factories Belle II and BESIII, the LHCb and PANDA experiments and
at a future Super-$c$-$\tau$ Factory.
Precise lattice calculations provide valuable input
and several results have been obtained by simulating up, down and strange quarks in
the sea. We investigate the impact of a charm quark in the sea on
the charmonium spectrum, the renormalization group invariant charm-quark mass $\Mc$
and the scalar charm-quark content of charmonium.
The latter is obtained by the direct computation of the mass-derivatives of the
charmonium masses.
We do this investigation in a model, QCD with two degenerate charm quarks.
The absence of light quarks allows us to reach very small lattice spacings
down to $0.023~$fm.
By comparing to pure gauge theory
we find that charm quarks in the sea affect the hyperfine splitting at a
level below 2\%.
The most significant effects are 5\% in $M_c$ and 3\% in the value of the charm quark content of the $\eta_c$ meson.
Given that we simulate two charm quarks these estimates are upper bounds
for the contribution of a single charm quark.
We show that lattice spacings $<0.06~$fm are needed for safe continuum extrapolations
of the charmonium spectrum with O($a$) improved Wilson quarks.
A useful relation for the projection to the desired parity of operators in two-point
functions computed with twisted mass fermions is proven.
\end{abstract}

\maketitle

\tableofcontents

\section{Introduction}\label{s:introduction}
\newcommand{\mscale}{{\cal S}}

The charmonium system is frequently characterized as the ``hydrogen atom'' of
meson spectroscopy owing to the fact that it is non-relativistic enough to be
reasonably well described by certain potential models \cite{Eichten:1974af}.
It is a perfect testing ground
for a comparison of theory with experiment. Over the last years, there has been a
renewed interest in spectral calculations with charmonia because of the experimental
discovery of many states which are not predicted by potential models \cite{Olsen:2015zcy},
e.g. the so-called X,Y,Z states, like the $X(3872)$ state \cite{Choi:2003ue} or the
$P_c$ pentaquark candidates \cite{Aaij:2015tga}.
More exciting experimental data are expected from the
B-factories Belle II \cite{Kou:2018nap} and BESIII \cite{BESIII:2016adj},
the LHCb experiment \cite{Bediaga:2018lhg},
the PANDA experiment at FAIR \cite{PANDA:2018zjt}
and at a future Super-$c$-$\tau$ Factory \cite{Eidelman:2015wja}.

Simulations of QCD on the lattice are a first-principle tool for precision
computations of charmonium states below the open charm thresholds ($D\bar{D}$ etc.)
\cite{Liu:2012ze,Cheung:2016bym,Follana:2006rc,DeTar:2018uko,Padmanath:2018tuc,Kalinowski:2015bwa},
see also \cite{Knechtli:2019bqx}.
States above the
open charm thresholds decay strongly and multi-hadron channels need to be included
for a full treatment. The masses of these resonances can be computed
in the approximation that they are treated as stable
and are accurate up to the hadronic width \cite{Liu:2012ze,Cheung:2016bym}.

For the computation of the charmonium spectrum the relevant quarks to include in the
lattice simulations are $u$, $d$, $s$, and $c$.
The question which we address in this work is the necessity to include the charm quark
$c$ in the sea, i.e. as a dynamical quark which contributes through loops and not
only as a valence quark.
QCD with $\nf=2+1\;(u,d,s)$ dynamical quarks is cheaper to simulate than QCD with
$\nf=2+1+1\;(u,d,s,c)$ dynamical quarks.
Adding a dynamical charm quark requires finer lattices than they are needed
for the lighter quarks and complicates the tuning of the parameters.

For processes at energies $E$ which are much smaller than the charm-quark mass $\Mc$
the charm quark decouples \cite{Weinberg:1980wa}.
It can be integrated out
and its effects are absorbed in the renormalization of the gauge coupling and
light quark masses, and in small corrections proportional to inverse powers of
$\Mc$.
In \cite{Bruno:2014ufa,Knechtli:2017xgy,Athenodorou:2018wpk} the
decoupling of the charm quark at low energies was studied in a model, namely
QCD with $\nf=2$ degenerate heavy quarks of mass $1.2\,\Mc \gtrsim M \gtrsim \Mc/8$.
Simulations at very small lattice spacings down to $a=0.023\,\fm$ in physical
volumes comparable to those used in the Yang--Mills theory allow to
control the continuum limit.
Concerning the renormalization the model study confirmed that treating decoupling
in perturbation theory only introduces small non-perturbative corrections which
can be estimated through the model calculation. Denoting
a low energy scale of mass dimension one by $\mscale$,
the mass-scaling function defined by
\begin{eqnarray}\label{e:etargi}
  \etargi &=& \frac{M}{\mscale}\,\frac{\partial \mscale}{\partial M} \,,
\end{eqnarray}
where $M$ is the renormalization group invariant mass of the heavy quark,
is universal (i.e. it does not depend on the specific scale chosen)
up to non-perturbative $1/M^2$ corrections $\Delta \etargi_\mathrm{NP}$.
In \cite{Athenodorou:2018wpk} the conclusion was that
$\Delta \etargi_\mathrm{NP} < 0.014$ for the charm quark \emph{in QCD}.
We emphasize that \eq{e:etargi} corresponds to the charm quark content of the
nucleon and is needed to compute the cross-section of the scalar interaction of
dark matter with nucleons.

In this work we extend the model study of charm loop effects
to observables which explicitly depend on a valence charm quark.
The paper is organized as follows.
In \Sect{s:model} we introduce the model.
\Sect{s:simulations} explains our lattice setup based on twisted mass
fermions at maximal twist, the observables and the computation
of their derivative with respect to the quark mass.
In \Sect{s:results} we present our results for the charm loop effects, specifically
in the charmonium spectrum and the renormalized charm-quark mass.
We also compute the generalization of the mass-scaling function in \eq{e:etargi}
to describe the charm-quark mass-dependence of the charmonium states.
All our results are evaluated after continuum extrapolation and we discuss
the size of lattice artifacts.
\Sect{s:conclusions} contains the summary of this work.
Appendix~\ref{s:parity} shows how to construct two-point functions which project
to definite parity states with twisted mass fermions.
In Appendix~\ref{sec:tables} the charmonium masses obtained on our ensembles are listed.

\section{Model}\label{s:model}
\newcommand{\ev}[1]{\left\langle #1 \right\rangle}

Consider QCD with quarks $q^i$, $i=\{u,d,s,c\}$. We denote their Dirac operators
by $D_i$.
Our goal is to estimate the contribution of charm-quark loops in physical
observables $A[q^i,U]$, where $U$ represents the gauge field.
The expectation value of the observable is
\begin{eqnarray}
  \ev{A[q^i,U]} & = &
  \frac{1}{Z}\int \mathcal{D}[U]\;
  \left(\prod_{j=u,d,s}\det D_j\right)\; \det D_c\;
  \tilde{A}[D^{-1}_i,U]\; {\rm e}^{-S[U]}
\end{eqnarray}
The charm-quark loop effects\footnote{
  Notice that here we mean non-perturbative effects due to quark loops
  on arbitrary gauge backgrounds.}
stem from the determinant $\det D_c$.
Quenching the charm, i.e. setting $\det D_c=1$ means neglecting the charm loops.
This approximation is made in the computations of the charmonium spectrum of Refs.~
\cite{Liu:2012ze,Cheung:2016bym,Follana:2006rc,DeTar:2018uko,Padmanath:2018tuc}.
In order to assess how good this approximation is, one would need a comparison
in the continuum limit with
simulations where a dynamical charm quark is added. Assuming this was possible,
the comparison would be superfluous since one would stick with the more
complete theory anyhow.
But adding a dynamical charm quark means a significant increase in the complexity
and costs of the simulations. This is so because of the additional 
tuning of the charm quark mass and the combination of small lattice spacings, which are
required by the large charm-quark mass and the large physical volumes, which are needed
to accommodate the light mesons. So the really interesting
question is if it is possible to decide whether
a dynamical charm quark is necessary \emph{before} doing the simulations.

This is why the study of a model, QCD with just $\nf=2$ degenerate charm quarks,
is appealing. 
Observables in this model are defined in terms of a doublet of charm quarks
$q^c=(c_1~,c_2)$ and their expectation value is
\begin{eqnarray}
  \ev{A[q^{c},U]} & = &
  \frac{1}{Z}\int \mathcal{D}[U]\;
  (\det D_c)^2\;
  \tilde{A}[D^{-1}_{c},U]\; {\rm e}^{-S[U]} \\
  & \equiv & \left\langle \tilde{A}[D^{-1}_{c},U] \right\rangle^{\rm gauge}\,.
\end{eqnarray}  
After matching this theory with a Yang--Mills (or pure gauge) theory,
the difference in physical observable will be a direct measure of the effects of
charm-quark loops. There are two differences with respect to a comparison between
QCD with four ($u$, $d$, $s$, and $c$) and three ($u$, $d$, $s$) quarks: in the model
we miss the effects of the light quarks and we double the number of sea charm quarks.
Since what we are interested in is a comparison of a theory with and without
charm quarks in the sea we do not expect the light quarks to affect the
\emph{difference} of the same quantity computed in the two theories much.
The extra charm quark in the sea will make the effects larger. For a low energy
quantity, where the theory of decoupling applies, the effects scale proportionally
to the number of quarks \cite{Athenodorou:2018wpk}, so they are overestimated by
a factor of two in the model. For a quantity with a valence charm quark decoupling
does not apply in the obvious way\footnote{
  Decoupling might apply for differences of masses or for binding energies.}
and we consider the effects computed with two
charm quarks in the sea as an upper bound for those with only one charm quark.

\section{Simulations}\label{s:simulations}
In this section we introduce the lattice setup used for this work and all the observables under investigation. 
We mainly focus on quantities with an explicit charm-quark dependence, like charmonium masses, the hyperfine splitting 
and the renormalization group invariant quark mass.

\subsection{Actions and algorithms}

We use relatively simple and theoretically well understood lattice actions for our simulations. 
For the $\Nf=0$ ensembles the standard Wilson plaquette action~\cite{Wilson:1974sk} is employed. 
In the $\Nf=2$ case, a doublet of twisted mass Wilson fermions is added~\cite{Frezzotti:2000nk,Frezzotti:2003ni}. 
In massless schemes, theories with standard and with twisted-mass fermions share the same renormalization factors, 
as long as other details of the action are the same. We therefore also include a clover 
term~\cite{Sheikholeslami:1985ij} in our action. Although not necessary for $O(a)$ improvement of 
physical quantities~\cite{Frezzotti:2003ni} (at maximal twist), it has been shown to reduce $O(a^2)$ artifacts 
in some cases~\cite{Dimopoulos:2009es}, and more importantly gives us access to the wide range of renormalization 
factors that have been determined non-perturbatively in the past. In particular we benefit from the knowledge of 
the critical mass $\mcr$~\cite{Fritzsch:2012wq,Fritzsch:2015eka} and the axial current and pseudoscalar density 
renormalization factors $Z_A$~\cite{Luscher:1996jn,DellaMorte:2005xgj,DallaBrida:2018tpn} and $Z_P$~\cite{Juttner:2004tb,Fritzsch:2012wq}.

Since one of our goals is a detailed understanding of charm related lattice artifacts, we simulate also at
very fine lattice spacings, much finer than what is currently feasible in simulations that include light quarks.
Problems related to deficient sampling of topological sectors are avoided by the implementation of open boundary 
conditions in the time directions~\cite{Luscher:2011kk}. The spatial dimensions are kept periodic.

To summarize, our action is $S=S_{\rm g} + S_{\rm f}$, with gauge action
\begin{equation}
   S_{\rm g} = \frac{1}{g_0^2} \sum_p w(p) {\rm tr}\left[1 - U(p) \right]\, ,
\end{equation}
where the summation is over all oriented plaquettes $p$ on the lattice, weighted by $w(p)$ which is
one everywhere except for spatial plaquettes on the temporal boundary time-slices, where it is $1/2$.
$U(p)$ is the product of four SU(3) gauge fields $U_\mu(x)$ around the elementary plaquette $p$.
Gauge fields are periodic in spatial directions and absent on temporal links sticking out of the lattice
(i.e. open boundaries).
The free parameter of the gauge action is the bare coupling $g_0^2 \equiv 6/\beta$.
In case of the $\Nf=2$ simulations, a fermionic action is added
\begin{equation}
   S_{\rm f} = \sum_x a^4 \, \bar \chi(x) [D\chi](x)\, ,
\end{equation}
where $\chi = (c_1,c_2)^\top$ is a flavor doublet of quarks and the Dirac operator is
\begin{equation}
   D = D_{\rm w} + D_{\rm sw} + m_0 + i\mu \gamma_5 \tau^3\, ,
\end{equation}
with bare mass $m_0$ and twisted bare mass $\mu$. The third Pauli matrix $\tau^3$ in the 
twisted mass term acts in flavor-space, all other terms of the operator are flavor
diagonal.
\begin{equation}
   D_{\rm w} = \sum_{\mu=0}^3 \frac{1}{2}\left( \gamma_\mu \left[\nabla^*_\mu+\nabla_\mu\right] - \nabla^*_\mu\nabla_\mu \right)
\end{equation}
is the massless Wilson operator, containing the usual covariant forward and backward finite difference
operators $\nabla_\mu\chi(x) = U_\mu(x)\chi(x+\hat\mu)-\chi(x)$ and 
$\nabla^*_\mu\chi(x) = \chi(x) - U^\dagger_\mu(x-\hat\mu)\chi(x-\hat\mu)$. Finally, the operator
in the Sheikholeslami-Wohlert term acts as
\begin{equation}
   D_{\rm sw}\chi(x) = c_{\rm sw} \sum_{\mu,\nu=0}^3 \frac{i}{4}\sigma_{\mu\nu}\hat F_{\mu\nu}(x)\chi(x)\, .
\end{equation}
A symmetric discretization of the field strength tensor $\hat F_{\mu\nu}$, as e.g. in~\cite{Luscher:1998pe}, is used. The fermionic fields are periodic in spatial directions and satisfy $D\chi=0$ on the first and last time-slice of the lattice. The fermionic part of the action has dimensionless simulation parameters $\kappa \equiv \frac{1}{2am_0+8}, a\mu$ and $c_{\rm sw}$. The above choice for the actions corresponds to setting the gluonic and fermionic boundary improvement terms to their tree level values.

Both $\Nf=0$ and $\Nf=2$ theories are simulated with a Hybrid Monte Carlo (HMC)~\cite{Duane:1987de} algorithm. The molecular dynamics
equations are integrated using a fourth order Omelyan-Mryglod-Folk integrator. In the case with fermions a multi-level variant is employed, with fermionic forces being integrated with a coarser step size than the forces deriving from the gauge action. In addition the quark determinant is factorized into two factors which are then represented by two separate path integrals over pseudo fermion fields~\cite{Hasenbusch:2001ne}.

The costs are dominated by solutions of the Dirac equation. The relatively high quark masses in our simulations, mean that a standard conjugate gradient algorithm is often more efficient than more complicated preconditioned variants. On the finer lattices however, SAP preconditioning~\cite{Luscher:2003qa} of the equations involving the light Hasenbusch mass is beneficial. Our simulations are carried out using a variant of {\tt openQCD}~\cite{Luscher:2012av}. A minor change allows us to choose a different twisted mass parameter in the SAP preconditioner than 
in the simulation \cite{Athenodorou:2018wpk,Alexandrou:2016izb}.
\Tab{tab:ensembles}~summarizes our simulation parameters.

\begin{table}[ht]
\begin{center}
{\footnotesize
\renewcommand{\arraystretch}{1.4}
\renewcommand{\tabcolsep}{5.5pt}
\begin{tabular}{c | c c c c c c c c}
\toprule
$N_f$ & ID & $\frac{T}{a}\times\left(\frac{L}{a}\right)^3$ &  $\beta$  & $\kappa$    & $a \mu$            & $\sqrt{t_0}m_{P}$ & $t_0/a^2$ &  MDUs\\
\midrule
2 & E &$95\times 24^3$  &  5.300  & 0.135943    & 0.36151  & 1.79303(55)      & 1.23907(82) & 8000\\
 & N & $119\times 32^3$                              &  5.500  & 0.136638    &  0.165997          & 1.8048(15)      & 4.4730(93) & 8000\\
 & O & $191\times 48^3$  &  5.600  & 0.136710    & 0.130949  & 1.7656(14) & 6.561(12) & 8000\\
 & P & $119\times 32^3$                              &  5.700  & 0.136698    & 0.113200           & 1.7931(28)     & 9.105(35) & 17184\\
 & S & $191\times48^3$                               &  5.880  & 0.136509    & 0.087626           & 1.8130(29)     & 15.621(60) & 23088\\
 & W & $191\times 48^3$                              &  6.000  & 0.136335    & 0.072557           & 1.8075(43)     &22.39(12) &  22400\\
\midrule
0 & qN & $119\times 32^3$                               &  6.100  &    --       &    --              & --  & 4.4329(38) & 64000 \\
 & qP & $119\times 32^3$                               & 6.340  &    --       &    --              & --  & 9.037(30) & 20080\\
 & qW & $191\times 48^3$                               &  6.672  &    --       &    --              & --  & 21.925(83) & 73920\\
 & qX & $191\times 64^3$                               & 6.900  &    --       &    --              & --  & 39.41(14) & 160200\\
\bottomrule
\end{tabular}
}
\end{center}
\caption[Simulation parameters of our ensembles.]{Simulation parameters of our ensembles. The columns show the lattice sizes, 
the gauge coupling $\beta=6/{g_0^2}$, the critical hopping parameter, the twisted mass parameter $\mu$, the pseudoscalar mass in $t_0$ units, 
the hadronic scale $t_0/a^2$ defined in \cite{Luscher:2010iy} and the total statistics in molecular dynamics units. Note that even though
the number of sites in the temporal direction is even, the temporal extent $T$ is an odd multiple of $a$ due to the open boundaries. The links
pointing out of the lattice volume are absent.
}\label{tab:ensembles}
\end{table}

The simulation algorithm performs very well. In particular no increased critical slowing down due to deficient sampling of topological sectors can be observed. The scaling of the exponential auto-correlation time with the lattice spacing is compatible with the expected $\tau_{\rm exp} \propto a^{-2}$ behavior~\cite{Luscher:2011kk}. The expected scaling of autocorrelation times with open boundary conditions has been shown in
Figure 8 of Ref.~\cite{Athenodorou:2018wpk}.

\subsection{Observables}

\subsubsection{$t_0$}
The Wilson-flow equation \cite{Luscher:2010iy,Narayanan:2006rf}
\begin{equation}
   \frac{\partial V_\mu(x,t)}{\partial t} = -g_0^2 \left(\partial_{x,\mu} S_{\rm g}[V] \right) V_\mu(x,t)
   ,\qquad V_\mu(x,0) = U_\mu(x)\, ,
\end{equation}
relates a ``smeared'' gauge field $V_\mu(x,t)$ at flow time $t$ to 
the original gauge field $U_\mu(x)$, that is integrated over in the path integral.
$S_{\rm g}[V]$ is a gauge action of the smeared fields, in our case the Wilson
plaquette action, and the link differential operator $\partial_{x,\mu}$ is
defined in the usual way~\cite{Luscher:2009eq,Luscher:2010iy}.
It has been shown that correlators constructed from gauge fields at $t>0$ are
automatically renormalized \cite{Luscher:2011bx}. Among other things, this allows to define
the low-energy length scale $t_0$~\cite{Luscher:2010iy} as the flow time $t$ at which 
\begin{equation}
   t^2 \langle E(t) \rangle = 0.3\, .
\end{equation}
In this equation $E(t)$ denotes the Yang-Mills action density at flow-time $t$,
away from the temporal boundaries. A different discretization than the one 
used in the simulations may be used. We follow~\cite{Luscher:1998pe} and use a symmetrized clover definition
\begin{eqnarray}
    E(x,t) &=& \frac{1}{4}G_{\mu\nu}^a G_{\mu\nu}^a\, ,
\end{eqnarray}
where $G_{\mu\nu}^a(x,t)$ are the Lie algebra components of the lattice field
strength tensor.

\subsubsection{Isovector meson masses}
We study mesons that are ground states in the channels that are excited by 
operators $\bar \psi \Gamma \tau^a \psi$. Twisted mass fermions at 
maximal twist, $\bar\chi$ and $\chi$,  are related to the fields in the 
physical basis by
\begin{eqnarray}
   \psi      &=& \frac{1+i\gamma_5\tau^3}{\sqrt{2}}\chi \, , \\
   \bar \psi &=& \bar\chi \frac{1+i\gamma_5\tau^3}{\sqrt{2}}\, .
\end{eqnarray}
This means that some operators take an unusual form. For flavor components
$\tau^1$ and $\tau^{2}$, the relations are summarized 
in Table~\ref{tab:meson_states}.

\begin{table}[ht]
\begin{center}
\renewcommand{\arraystretch}{1.4}
\renewcommand{\tabcolsep}{8pt}
\begin{tabular}{lllll}
\hline
State        & $J^{PC}$ & Particle & Physical basis & Twisted basis               \\
\hline
Scalar       & $0^{++}$ & $\chi_{c0}$ &$S^{1,2}=\bar{\psi}\frac{\tau^{1,2}}{2}\psi$ &  $\bar{\chi}\frac{\tau^{1,2}}{2}\chi$  \\
Pseudoscalar & $0^{-+}$ & $\eta_c$ &$P^{1,2}=\bar{\psi}\gamma_5\frac{\tau^{1,2}}{2}\psi$ &  $\bar{\chi}\gamma_5\frac{\tau^{1,2}}{2}\chi$  \\
Vector       & $1^{--}$ & $J/\psi$ &$V_{i}^{1,2}=\bar{\psi}\gamma_{i}\frac{\tau^{1,2}}{2}\psi$  &  $\pm\bar{\chi}\gamma_{i}\gamma_{5}\frac{\tau^{2,1}}{2}\chi$   \\
Axial vector & $1^{++}$ & $\chi_{c1}$ &$A_{i}^{1,2}=\bar{\psi}\gamma_{i}\gamma_{5}\frac{\tau^{1,2}}{2}\psi$  & $\pm\bar{\chi}\gamma_{i}\frac{\tau^{2,1}}{2}\chi$ \\
Tensor\footnote{The notation refers to the $\gamma$-structure of the operator.}
  & $1^{+-}$ & $h_c$&$T_{ij}^{1,2}=\bar{\psi}\gamma_i\gamma_j\frac{\tau^{1,2}}{2}\psi$  & $\bar{\chi}\gamma_i\gamma_j\frac{\tau^{1,2}}{2}\chi$\\
\hline
\end{tabular}
\end{center}
\caption{Typical interpolators for meson states and relations between physical and twisted basis. The particle name is the closest relative in nature.}\label{tab:meson_states}
\end{table}

Meson masses can be extracted from zero momentum correlation functions of the form
\begin{equation}\label{eq:mesoncorr}
   f_{\Op_1\Op_2}(x_0,y_0) = a^6 \sum_{\mathbf{x},\mathbf{y}} \langle \Op_1(x) \Op^\dagger_2(y) \rangle \, ,
\end{equation}
with various choices of the operators $\Op_i$. 
We work with definite flavor assignments, e.g.
$P^+ \equiv P^1 +iP^2 = \bar c_1 \gamma_5 c_2$.
Then, integrating over the fermions leaves us with a single connected 
diagram of the form 
$-\sum_{\mathbf{x},\mathbf{y}} \langle {\rm tr}[ \Gamma_1 \Prop_1(x,y) \Gamma_2 \Prop_2(y,x)]\rangle^{\rm gauge}$,
where $\Gamma_i$ are $4\times 4$ matrices related to the operators in the correlation function, 
and
$\Prop_1$ ($\Prop_2$) is the inverse Dirac operator with positive (negative) twisted mass term.
Spatial translation invariance could be exploited to eliminate one of the sums, 
which would allow to compute a correlator at the cost of 12 solutions of the Dirac
equation per 
choice of $y_0$. The signal however is highly improved, by keeping the two sums.
The trace can then be efficiently estimated stochastically. We use 
time-dilution with 16 $U(1)$ noise sources per time-slice, which amounts to 
16 inversions per $y_0$ value and Dirac structure.

An improved signal and exact symmetries are achieved by defining the averages
\begin{eqnarray}
	f_P(x_0-a) &\equiv& \frac{1}{2} \left( f_{PP}(x_0,a) + f_{PP}(T-x_0,T-a)  \right)\, , \label{eq:avcorr1} \\
   f_A(x_0-a) &\equiv& \frac{1}{2} \left( f_{PA}(x_0,a) - f_{PA}(T-x_0,T-a)  \right)\, , \\
   f_V(x_0-a) &\equiv& \frac{1}{6} \sum_{k=1}^3  \left( f_{V_kV_k}(x_0,a) + f_{V_kV_k}(T-x_0,T-a)  \right) \, , \\
   f_S(x_0-a) &\equiv& \frac{1}{2} \left( f_{SS}(x_0,a) + f_{SS}(T-x_0,T-a)  \right)\, , \\
   f_T(x_0-a) &\equiv& \frac{1}{6} \sum_{j>i} \left( f_{T_{ij}T_{ij}}(x_0,a) + f_{T_{ij}T_{ij}}(T-x_0,T-a)  \right).\label{eq:avcorr5}
\end{eqnarray}
Enforcing the continuum time reflection symmetries prevents opposite parity operators from
mixing, as explained in Appendix~\ref{s:parity}.
From the exponential decay of these correlators at $x_0 \gg a $, meson masses are extracted. 
First, effective masses are computed
\begin{equation}
   a m^{\rm eff}(x_0+a/2) \equiv \ln\left(\frac{f(x_0)}{f(x_0+a)} \right)\, ,\label{eq:effmass}
\end{equation}
and the meson mass is then given as a weighted plateau average 
\begin{equation}
	m = \frac{\sum\limits_{x_0 = t_{\rm low}}^{t_{\rm high}} w(x_0+a/2) m^{\rm eff}(x_0+a/2)}{\sum\limits_{x_0 = t_{\rm low}}^{t_{\rm high}} w(x_0+a/2)}\, .
\label{eq:plateau}
\end{equation}
The start of the plateau, $t_{\rm low}$, is chosen such that excited state contributions are completely negligible, and the 
weights $w$ are given by the inverse squared errors of the corresponding effective masses. All masses that we extract 
are those of iso-vector mesons. In the light sector these would be called pions or kaons 
($f_P,f_A$), $\rho$- or $K^*$-mesons ($f_V$), $a_0$, $f_0$, $K_0^{\star}$ ($f_S$), $h_1$, $b_1$ ($f_T$). 
However, since both our quarks have the mass of a charm-quark, the meson masses that we obtain are more 
comparable to the charmonia masses $\eta_c$, $J/\psi$,$\ldots$ respectively. The difference being, 
that these are iso-scalars and the determination of their masses would require the computation 
of disconnected (charm annihilation) diagrams.

\subsubsection{PCAC mass}
Partial conservation of the axial current is an operator relation
\begin{equation}
   \partial_\mu \hat A_\mu = 2 m_{\rm PCAC} \hat P\, .
\end{equation}
On the lattice it holds up to lattice artifacts, when inserted into any
correlation function, as long as $A$ and $P$ are at a different positions than
all other operators in the correlator. These lattice artifacts depend on the 
exact choice of correlation function, and can be quite large. We extract the
bare PCAC quark mass from
\begin{equation}\label{eq:mPCAC}
   \mPCAC = \frac{\tilde \partial_0 f_A}{2 f_P} \, ,
\end{equation}
where $\tilde \partial_\mu$ denotes the symmetric finite difference operator.
The lattice artifacts in this quantity increase, when the correlators are 
evaluated close to the boundary (small $x_0$). We form an average value from the 
time-slices in the plateau region away from boundaries. 

\begin{figure}[h]
   \centering
   \includegraphics[width=0.8\linewidth]{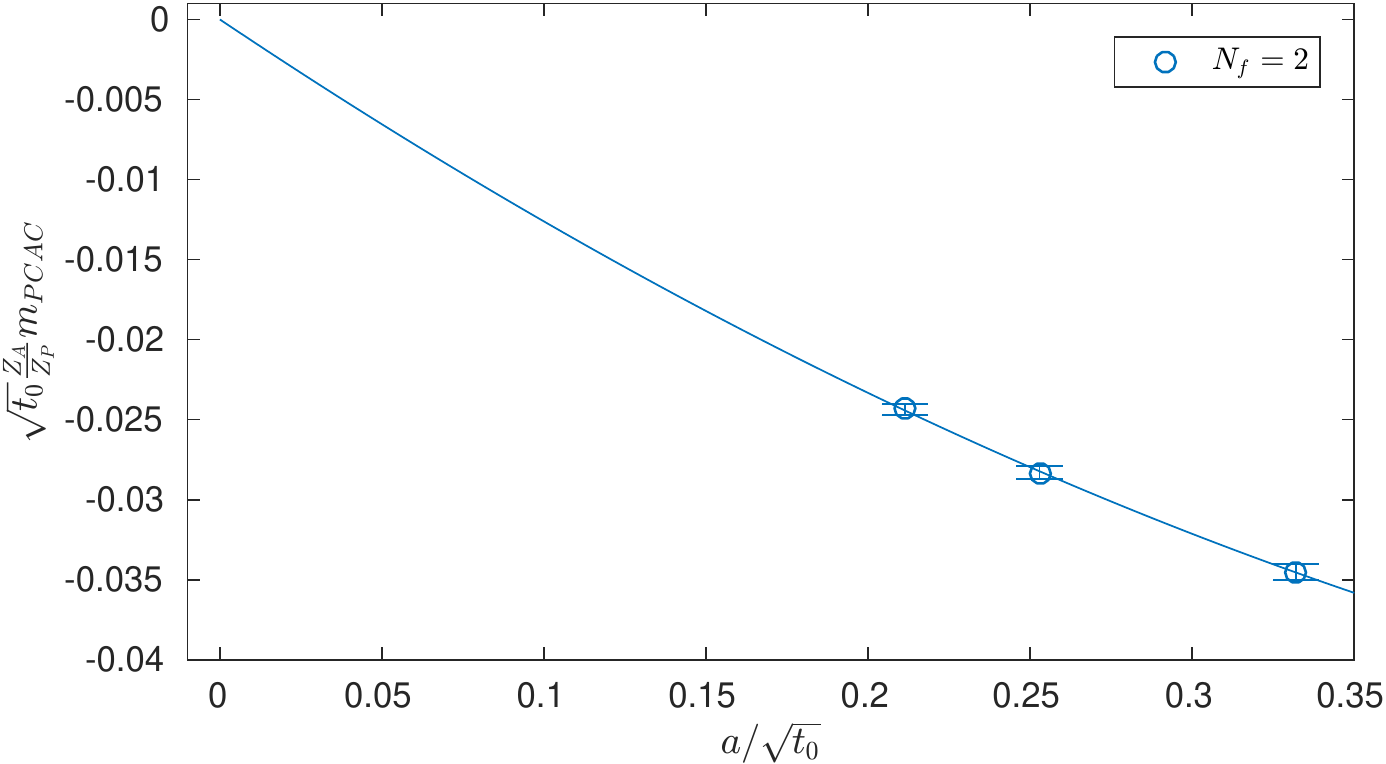}
   \caption{The standard mass contribution to the renormalized quark mass, which 
            vanishes in the continuum limit.}\label{fig:mPCAC}
\end{figure}

For us the main use of the PCAC mass is to find the critical value of the bare mass $m_0$,
i.e. the maximal twist condition.
It is given by the value at which $m_{\rm PCAC}=0$. Instead of determining 
it ourselves, we use very precise critical masses obtained in~\cite{Fritzsch:2012wq,Fritzsch:2015eka}. These
were computed from slightly different correlation functions in a finite volume,
and differ from ours by an $O(a)$ lattice artifact. We thus do not expect
the PCAC masses that we determine to be zero, but to be small and to vanish when
the continuum limit is approached. By computing them, we put this expectation 
to a test. \Fig{fig:mPCAC} demonstrates that indeed, up to lattice artifacts, we are at maximal twist. If we consider the usual definition of the twist angle 
\begin{equation}
\omega={\rm arctan}\left(\frac{\mu}{Z_A \mPCAC}\right),
\end{equation}
the largest deviation from maximal twist ($\omega=\pi/2$) that we encounter in our simulations is around $6\%$ in the ensemble E, whilst the smallest deviation is around $2\%$ in the ensemble W.

\subsubsection{RGI quark mass}
At maximal twist a renormalized quark mass is given by
$\overline m = Z_P^{-1} \mu$, and depends on the scale and scheme in which 
$Z_P$ was computed. Away from maximal the more general relation
\begin{equation}\label{eq:mbar}
   \overline m = Z_P^{-1} \sqrt{\mu^2 + Z_A^2 \mPCAC^2 } 
\end{equation}
holds. We neglect the (very small) contribution due to non-vanishing $\mPCAC$ in our determination, after veryfying that it is
compatible with being of $O(a)$. 
The axial current renormalization factor $Z_A$ is scale independent. It has been determined non-perturbatively in the $\Nf=0$ theory with our action, by exploiting current algebra relations in a massless Schr\"odinger functional~\cite{Luscher:1996jn}. The same technique has also been applied to the $\Nf=2$ theory~\cite{DellaMorte:2005xgj}. In this case also a more precise determination based on universal relations between correlators in a chirally rotated Schr\"odinger functional exists~\cite{DallaBrida:2018tpn}, and these are the values that we use here.
The pseudoscalar renormalization factor $Z_P$ depends on the renormalization scheme 
and scale. It is known non-perturbatively in the SF scheme 
in both $\Nf=0$~\cite{Juttner:2004tb} and $\Nf=2$~\cite{Fritzsch:2012wq} theories for a wide range of bare couplings, albeit at slightly different scales. The renormalized charm quark mass can thus be computed in the continuum limit, in this particular 
scheme. To be able to compare the two theories, also the scales should match.
We go one step further and compute directly the RGI masses, which are scale and
scheme independent. The necessary relations between renormalized and RGI masses
are well known for the scales and schemes used above, namely
$M/ \overline m = 1.157(12)$ in the $\Nf=0$ theory~\cite{Juttner:2004tb}, and $M/\overline m = 1.308(16)$ in the theory with two dynamical quarks~\cite{Fritzsch:2012wq}.

\subsubsection{Twisted mass derivatives}
We also computed the derivatives of all the observables above, with
respect to the twisted mass parameter $\mu$. The twisted mass derivative 
of a primary observable $A$ is given by
\begin{equation}\label{eq:dAdmu}
  \frac{{\rm d} \langle A \rangle}{{\rm d} \mu} =
  -\left\langle\frac{{\rm d}S}{{\rm d}\mu}A\right\rangle
   +\left\langle\frac{{\rm d}S}{{\rm d}\mu}\right\rangle\left\langle A\right\rangle 
   +\left\langle\frac{{\rm d}A}{{\rm d}\mu}\right\rangle\, .
\end{equation}

Most quantities we are interested in, are non-linear functions of various primary 
observables (e.g. $m_P$, which depends on the correlator $f_P$ at various distances in the 
plateau region). 
For these the chain rule dictates
\begin{equation}\label{eq:fder}
   \frac{{\rm d} f(\langle A_1\rangle,\ldots ,\langle A_N\rangle,\mu)}{{\rm d}\mu} = 
   \frac{\partial f}{\partial \mu} 
   + \sum_{i=1}^N \frac{\partial f}{\partial \langle A_i\rangle}\frac{{\rm d}\langle A_i\rangle}{{\rm d}\mu} .
\end{equation}

None of the observables that we consider have an explicit $\mu$ dependence,
so the last term in \eq{eq:dAdmu} is absent. 
The derivative of the action is  ${\rm d}S/{\rm d}\mu = \sum_x \bar \chi i\gamma_5\tau^3\chi$,
and this is all that is needed to compute the twisted-mass derivatives of 
purely gluonic observables. More precisely,
\begin{eqnarray}
\left\langle \frac{{\rm d}S}{{\rm d}\mu} A[U] \right\rangle 
&=& ia^4\sum_x \langle (\bar c_1(x) \gamma_5 c_1(x) - \bar c_2(x) \gamma_5 c_2(x)) A[U] \rangle \\
&=& ia^4\sum_x \left\langle {\rm tr}\left[\gamma_5\left({\Prop_1}(x,x)-\Prop_2(x,x)\right) \right]A[U]\right\rangle^{\rm gauge} \\
&=& -2\mu a^8\sum_{x,y}\left\langle {\rm tr}\left[\Prop_1^\dagger(x,y) \Prop_2(x,y) \right]A[U]\right\rangle^{\rm gauge}\, .
\end{eqnarray}
The last line is a consequence of the twisted-mass relation $D_1 - D_2 = 2i\gamma_5 \mu$
and allows for a more precise stochastic determination of the trace~\cite{Jansen:2008wv}.
We found that 64 $U(1)$ noise vectors are enough for the errors in the 
determination of the derivative to
be dominated by gauge-noise, rather than the noise from the stochastic trace
evaluation.

If the observables depend on fermionic
fields too, the first term of \eq{eq:dAdmu} gives rise to new contractions 
that have to 
be computed. These are different for every fermionic observable. In the case of
our two-point functions \eq{eq:mesoncorr} we find contractions of the form
$-a^{10}\sum_{\mathbf{x},\mathbf{y},z} \langle {\rm tr}[ \Gamma_1 \Prop_{2}(x,y) \Gamma_2 \Prop_1(y,x)]
{\rm tr}[\gamma_5(\Prop_1(z,z)-\Prop_2(z,z))]\rangle^{\rm gauge}$, 
that can be immediately computed because both traces have already been 
estimated for the evaluation of the correlator and of ${\rm d}S/{\rm d}\mu$ respectively, and new terms
\begin{equation}
ia^{10}\sum_{\mathbf{x},\mathbf{y},z}\left\langle
{\rm tr}
\left[
\gamma_5\Prop_2(z,y)\Gamma_2\Prop_1(y,x)\Gamma_1\Prop_2(x,z)
\right]-
{\rm tr}
\left[
\gamma_5\Prop_1(z,x)\Gamma_1\Prop_2(x,y)\Gamma_2\Prop_1(y,z)
\right]
\right\rangle^{\rm gauge}
\label{eq:new_contractions1}
\end{equation}
that require some attention. When evaluated stochastically together with the 
correlator itself, the number of necessary 
inversions is increased by a factor of 3. While the ${\rm d}S/{\rm d}\mu$ terms 
quantify the dependence on the sea-quarks, this last term gives the 
valance quark mass dependence of the correlator, which is generally much 
stronger - especially with heavy quarks. 

\subsubsection{Mass scaling functions}
At last, we also investigate the mass scaling functions
\begin{equation}\label{e:etargi_ccbar}
  \eta_x\equiv\frac{\mu}{m_x}\frac{{\rm d}m_x}{{\rm d}\mu} =
  \frac{M}{m_x}\frac{{\rm d}m_x}{{\rm d}M},
\end{equation}
where $m_x$ denotes the mass of a meson in a generic $x$ channel (scalar, pseudoscalar, vector, axial vector, tensor) 
and $M$ is the renormalization group invariant quark mass. Note that $\eta_x$ is a 
renormalized quantity and its continuum limit can be easily extracted from the measurements 
performed at different lattice spacings, without the need of any renormalization factor.
Notice that by the Hellman-Feynman-Theorem~\cite{Feynman:1939zza}, $\eta_x$ is proportional to the scalar charm
quark density between meson states $x$, i.e. the $\sigma$-term
\begin{equation}
   \eta_x = \frac{1}{m_x} \langle x | M_c (\bar c c)_{\rm RGI} | x \rangle\, .
\end{equation}

Once the twisted mass derivatives of the meson correlators are known, the 
determination of $\eta_x$ amounts to the evaluation of \eq{eq:fder} with a
particular function $f$. Since the action of $\Nf=0$ QCD does not depend on $\mu$,
the calculation is greatly simplified in this case. \Eq{eq:dAdmu} receives a single contribution of 
the form $\langle {\rm d}\tilde{A}[D^{-1},U]/{\rm d}\mu\rangle$. In the $\Nf=2$ theory on the other
hand,
also the $\mu$-derivative of the action must be taken into account.

\subsection{Parameters, tuning and mis-tuning corrections}\label{sec:tuning}
Apart from the lattice size, the bare parameters of the $\Nf=2$ simulations are 
the inverse bare coupling $\beta$, the bare mass $am_0$ and the bare twisted mass $a\mu$. The choice of $\beta$  corresponds to a choice of the lattice spacing.
We choose to simulate at $\beta\in\{5.3, 5.5, 5.6, 5.7, 5.88, 6.0\}$ which spans a wide range of lattice spacings, see \Tab{tab:a} and allows 
for very controlled continuum extrapolations.

The bare mass is set to its critical value $m_0=\mcr$. To achieve this, the values in~\cite{Fritzsch:2012wq} are fitted to a Pad\'e function, as described in~\cite{Athenodorou:2018wpk}. This puts us to maximal twist, up to $O(a^2)$. In this situation the physical quark mass
is given by the twisted mass parameter $\overline m = Z_P^{-1} \mu$. On our finest lattice at $\beta=6.0$ we choose
\begin{equation}\label{eq:amu}
   a\mu = \frac{M_c}{\LambdaMS} \times Z_P(L_1^{-1})\times \frac{\overline m_c(L_1)}{M_c}\times  \LambdaMS L_1 \times \frac{a}{L_1} \, ,
\end{equation}
where the ratio of the RGI charm quark mass and the two flavor $\Lambda$ parameter is set to 4.87, the pseudoscalar renormalization factor at scale $L_1^{-1}$ and $\beta=6.0$ in the SF scheme is $Z_P=0.5184(33)$~\cite{Fritzsch:2012wq}, the relation between a renormalized quark mass in the SF-scheme at scale $L_1^{-1}$ and 
the RGI quark mass $M$ is known in the continuum~$M/\overline{m}(L_1^{-1}) =1.308(16) $~\cite{Fritzsch:2012wq}, and 
$\LambdaMS L_1 = 0.649(45)$~\cite{DellaMorte:2004bc}. Finally $L_1$ in lattice units is obtained by an interpolation of $L_1/a$ vs $\beta$ data from~\cite{Fritzsch:2012wq} to $\beta=6.0$. A quadratic fit of ${\rm log}(L_1/a)$ as a function of $\beta$, describes the data very well and yields $L_1/a_{\beta=6.0} = 17.27(70)$.
The quite substantial errors mean, that our simulated mass corresponds to the charm quark mass only up 
to about $10\%$. This is however fully sufficient for us, as long as the relative mass differences between
the different ensembles are under better control. To achieve this, we do not use \eq{eq:amu} at the 
other lattice spacings. Instead we proceed as follows:
the dimensionless, renormalized quantity
\begin{equation}\label{e:tuning}
  \sqrt{t_0} m_P = \tmPstar
\end{equation}
is determined on the ensemble with
the finest lattice spacing. On 
the coarser lattices $a\mu$ is tuned such that the same value of $\sqrt{t_0} m_P$ is obtained. 
This condition determines the bare twisted mass parameter very precisely and ensures that all ensembles 
have the same renormalized quark mass up to $O(a^2)$. 
Finally, the clover coefficient $\csw$ is set to its non-perturbatively determined value~\cite{Jansen:1998mx}.

\begin{figure}
   \includegraphics[width=0.8\linewidth]{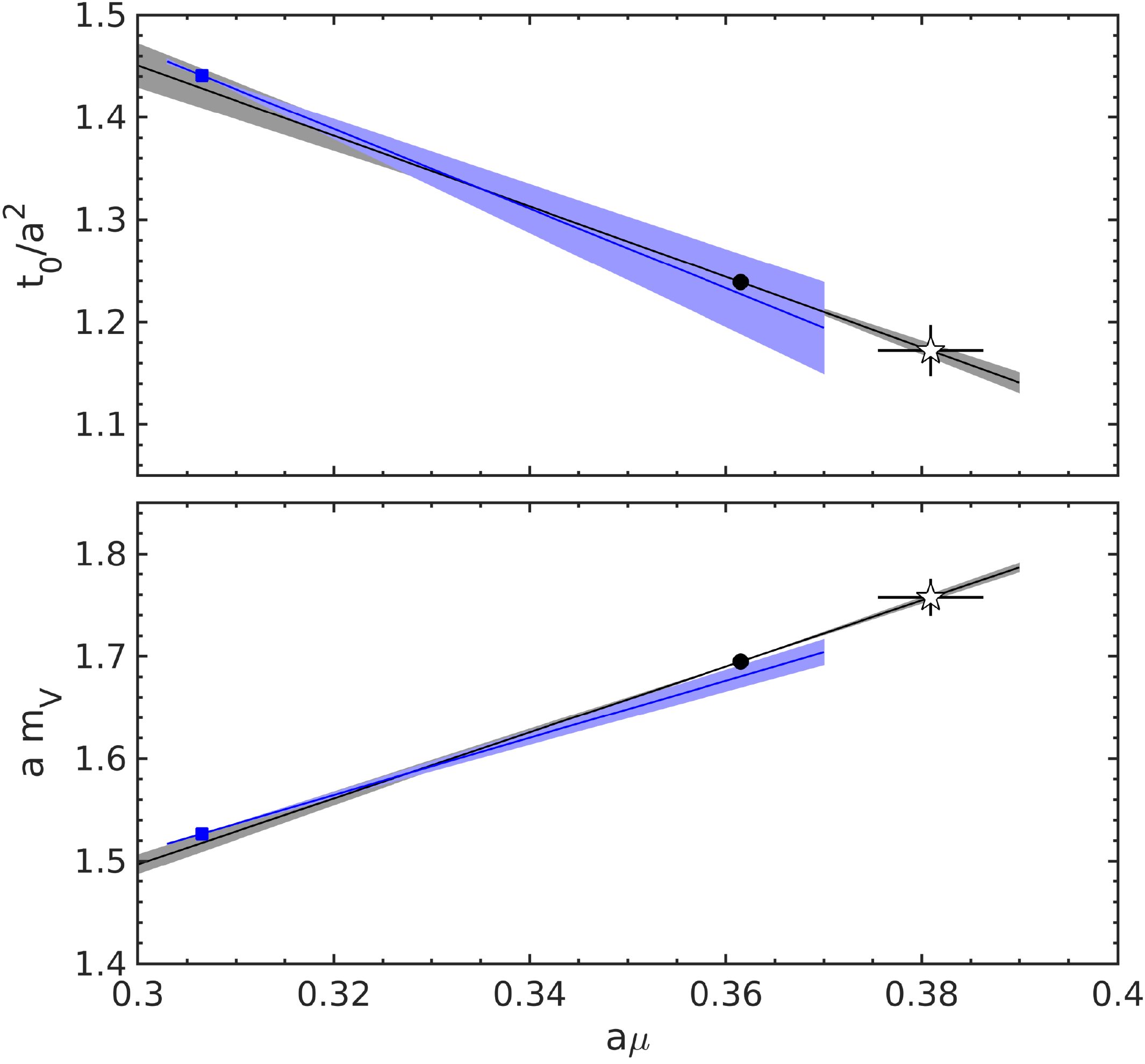}
   \caption{The solid square and circular markers are direct simulation results for
   $t_0/a^2$ (top) and $am_V$ (bottom) on our coarsest ensembles with
   $\beta=5.3$. The simulations were carried out at slightly different masses, 
   namely $a\mu=0.36151$ (circle) and $a\mu=0.30651$ (square). The lines, with their
   respective error bands illustrate the value and error of the derivative of the 
   observable with respect to the twisted mass parameter. The pentagram
   depicts the values obtained at the tuning point \eq{e:tuning}                 .
   Its vertical error bar is the complete error, including all correlations.}\label{fig:shifts}
\end{figure}

The tuning of the twisted mass parameter can be only carried out to a limited precision - at most to
within the statistical errors. To account for the mis-tuning, a correction is applied to all observables,
based on the computed twisted mass derivatives. First a target tuning point $\mu^\star$ is determined
\begin{equation}
   \mu^\star = \mu + \left(\sqrt{t_0}m_P - \tmPstar\right)\left(\frac{{\rm d}\sqrt{t_0}m_P}{{\rm d}\mu} \right)^{-1} \, ,
\end{equation}
and afterwards all quantities, denoted by $\Phi$ below, are corrected
\begin{equation}
   \Phi(\mu^\star) = \Phi(\mu) + (\mu^\star-\mu)\frac{{\rm d}\Phi}{{\rm d}\mu} \, .
\end{equation}
The error of the tuning point $\mu^*$ is propagated to the value of $\Phi(\mu^*)$ taking all correlations
into account.
It is assumed that the initial tuning was precise enough for the omitted 
quadratic terms to be negligible, compared to the statistical precision. 
\Fig{fig:shifts} demonstrates the procedure. 

A comparison with direct simulations indicates that even for large shifts of $\approx 15\%$ in $a\mu$ 
the linear approximation works well. The true shifts, that are needed are all much smaller, 
at most $5.40 \%$. Note that the $\mu$-shifts could also be computed using the 
mass reweighting, as explained in~\cite{Finkenrath:2013soa,Leder:2015fea}.

\begin{figure}[h]
\centering
\includegraphics[width=0.8\linewidth]{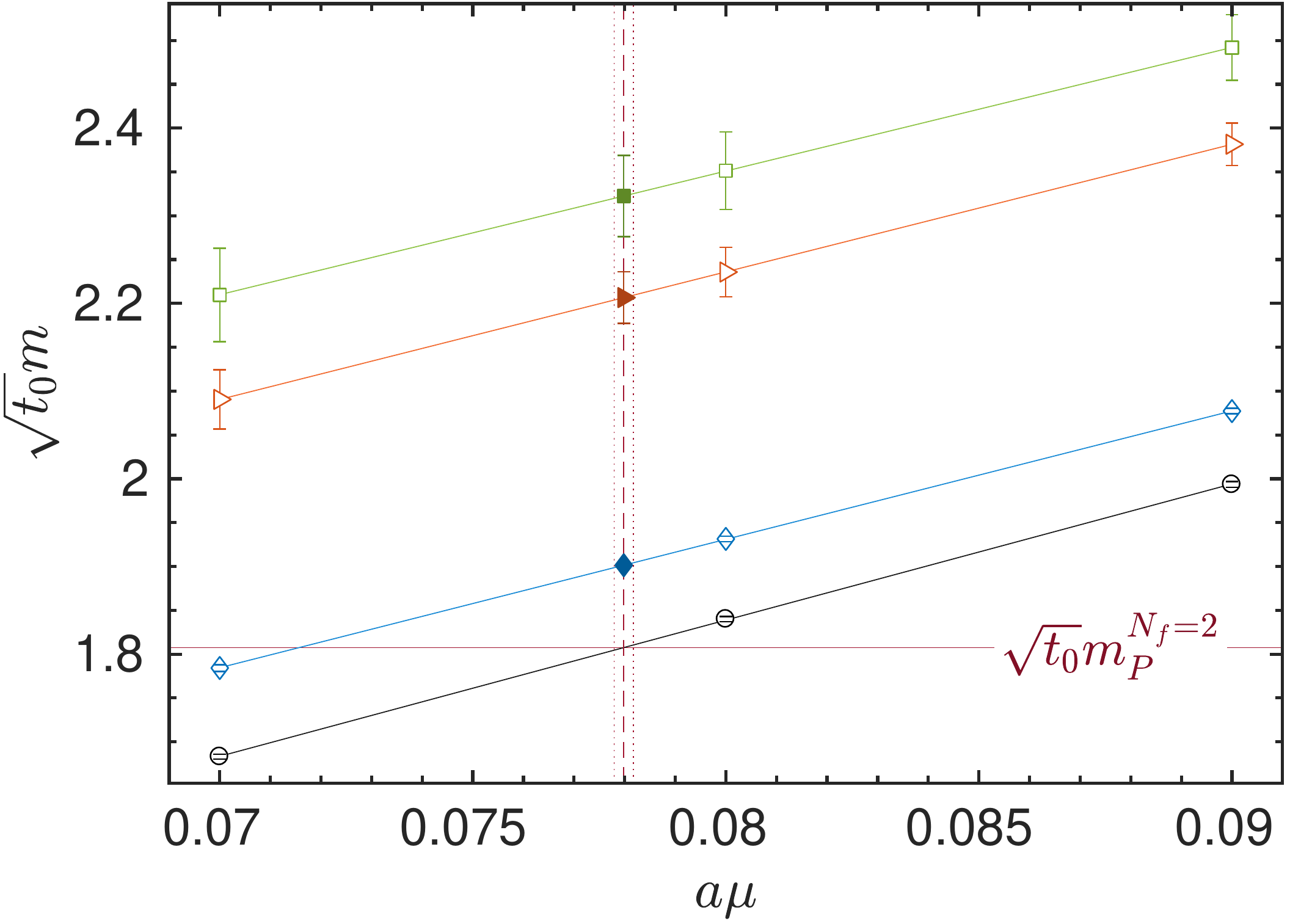}
\caption{Interpolation of the measured pseudoscalar masses (circles) on the $\Nf=0$ ensemble qW (see Table~\ref{tab:ensembles}). The horizontal line depicts the tuning point \eq{e:tuning}. The vertical lines are the
resulting interpolated twisted mass parameter $a\mu^{\star}$ and its statistical error. The measured vector, scalar and tensor mesons masses (diamonds, triangles and squares respectively) can then be interpolated to the tuning point, resulting in the corresponding solid markers. In their error bars all the correlations among the data have been taken into account.}\label{fig:shifts_quenched}
\end{figure}

The $\Nf=0$ simulations are carried out at $\beta\in\{6.1,6.34,6.672,6.9 \}$. 
The valence quarks have $m_0=\mcr$~\cite{Luscher:1996ug}, non-perturbative $\csw$ from~\cite{Luscher:1996ug} 
and three values of the twisted mass parameter, chosen such that a short interpolation 
to the value of $\sqrt{t_0}m_P$ given in \eq{e:tuning} can be performed. An example of 
this procedure is shown in Figure~\ref{fig:shifts_quenched}. Since decoupling applies 
to $t_0$, the condition \eq{e:tuning} means that the quark mass in the $\nf=0$ theory is 
the same as in the theory with two flavors, up to $O(a^2)$ and tiny $O(\Lambda^2/M_c^2)$ 
power corrections~\cite{Knechtli:2017xgy,Athenodorou:2018wpk}.

\subsection{Data Analysis}
We use the $\Gamma$-method~\cite{Wolff:2003sm} for the determination of statistical uncertainties. 
Observables like the effective mass~\Eq{eq:effmass} are non-linear functions of 
``primary observables'', and their errors are determined as described in~\cite{Schaefer:2010hu}. When incorporating the mis-tuning corrections of \Sect{sec:tuning} the necessary nonlinear 
functions can become quite unwieldy. For instance, the vector meson mass at $\mu^\star$ depends
on the vector correlator in the plateau region, but also on the pseudoscalar correlator in its 
plateau region, to determine how big a shift in $\mu$ is required. Furthermore, the 
vector mass depends on the $\mu$-derivatives of these correlators, on the $\mu$-derivative
of the action and on the $\mu$-derivative of the action times the correlators. 
Combinations like $\sqrt{t_0}m_V$ depend on even more primary data.

\section{Results}\label{s:results}
\subsection{Raw results}
\begin{figure}[h]
 \centering
 \includegraphics[width=0.8\linewidth]{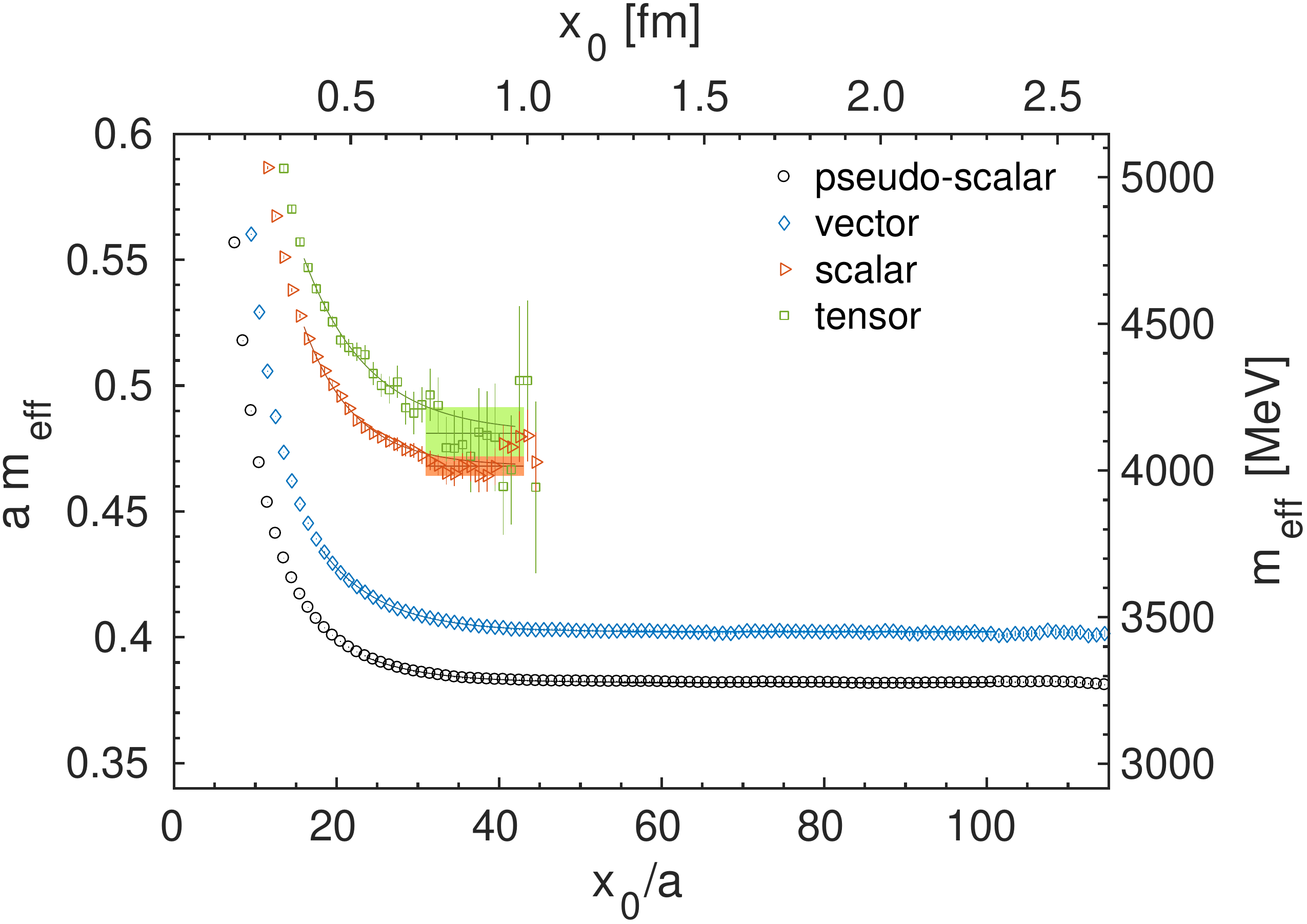}
 \caption{The effective masses for the pseudoscalar (circles), vector (diamonds),
 scalar (triangles) and tensor (squares) channels are displayed, together with 
 the plateau average and its error band. The fit to the leading correction~\eq{eq:meffcorr} 
 is also shown.}\label{fig:effmass}
\end{figure}

We measured all observables described in the previous section on all ensembles, 
except of $m_T$ and $m_S$ which were measured only on a subset and the 
mass derivatives, which were not measured on the $W$ ensemble.
A somewhat delicate issue is the proper choice of the plateau regions over which
the effective masses are averaged. The leading correction to a constant
effective mass is given by
\begin{equation}\label{eq:meffcorr}
   am^{\rm eff}(x_0+a/2) = am + c\ e^{-\Delta_1 x_0} + O(e^{-2\Delta_1 x_0}) 
         + O(e^{-\Delta_2 x_0})\, ,
\end{equation}
where $\Delta_1$ ($\Delta_2$) is the distance between $m$ and the first (second)
excited state. In a first preliminary fit we determine $\Delta_1$ and $c$. We
are then in the position to choose the plateau region such, that the influence of 
the excited states on the plateau average \eq{eq:plateau} is negligible compared
to its statistical uncertainty. The thus determined plateau regions are
collected in \Tab{tab:plateaus}. \Fig{fig:effmass} demonstrates the procedure 
for the case of ensemble $W$. The effective masses in the axial-vector channel
become too noisy, before a clean plateau is reached and are hence excluded from
the tables.

The results for the plateau averages are summarized in \Tab{tab:masses} in 
Appendix~\ref{sec:tables}, 
which shows the results at
the simulated parameters, as well as the values corrected for small mis-tunings 
in the twisted mass parameter.

\begin{table}[h]
   \centering
   \begin{tabular}{c c c c c}
   \toprule
   ID & $m_P$  & $m_V$    & $m_S$ & $m_T$ \\
   \midrule
   E  & 21-35   & 21-35   & -     & -     \\
   N  & 30-58   & 30-58   & 26-46 & 26-46 \\
   O  & 34-71   & 34-71   & -     & -     \\
   P  & 37-71   & 37-71   & 25-51 & 30-51 \\
   S  & 47-101  & 47-101  & -     & -     \\
   W  & 55-101  & 55-101  & 31-41 & 31-41 \\
   \midrule
   qN & 32-58   & 32-58   & 26-42 & 26-42 \\
   qP & 39-71   & 39-71   & 28-51 & 28-41 \\
   qW & 60-101  & 60-101  & 35-44 & 30-43 \\
   qX & 104-173 & 104-173 & -     & -     \\
   \bottomrule
   \end{tabular}
   \caption{The meson masses are determined from effective masses in the region
	$t_{\rm low} < x_0+a/2 < t_{\rm high}$. The table shows 
	$t_{\rm low}/a-t_{\rm high}/a$ for the different ensembles and channels.}\label{tab:plateaus}
\end{table}

\subsection{Continuum extrapolations}
We perform continuum extrapolations of dimensionless quantities. These are
either ratios of meson masses, namely $m_V/m_P$, $m_S/m_P$ and $m_T/m_P$, or the 
mass-scaling functions $\eta_P$ and $\eta_V$. One last quantity is the renormalized 
quark mass. We take the continuum limit of the dimensionless ratio of $\overline m$ and
$m_P$. All fits are restricted to a region where the data can be well 
described by the expected leading scaling violations of order $a^2$. This means,
neglecting data with lattice spacings coarser than $a^2/t_0 > 0.25$.

\begin{figure}[h]
   \centering
   \includegraphics[width=0.8\linewidth]{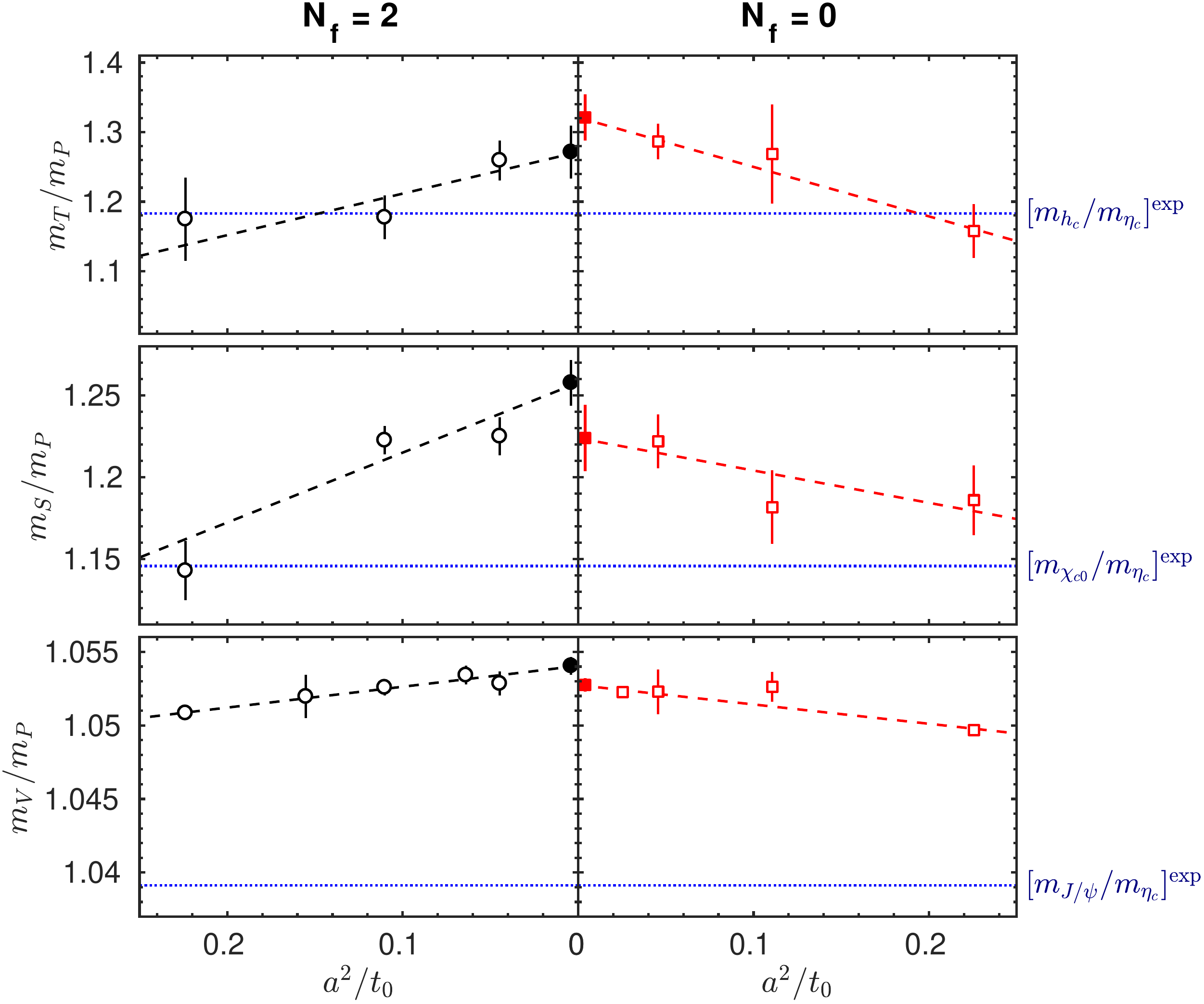}
   \caption{Continuum limits of the meson mass ratios $m_V/m_P, m_S/m_P$ and
   $m_T/m_P$ in both the $\Nf=2$ (left) and $N_f=0$ (right) theories. The dotted lines
   indicate the value of the corresponding ratio in nature.}\label{fig:spectrum}
\end{figure}

\begin{figure}[h]
   \centering
   \includegraphics[width=0.8\linewidth]{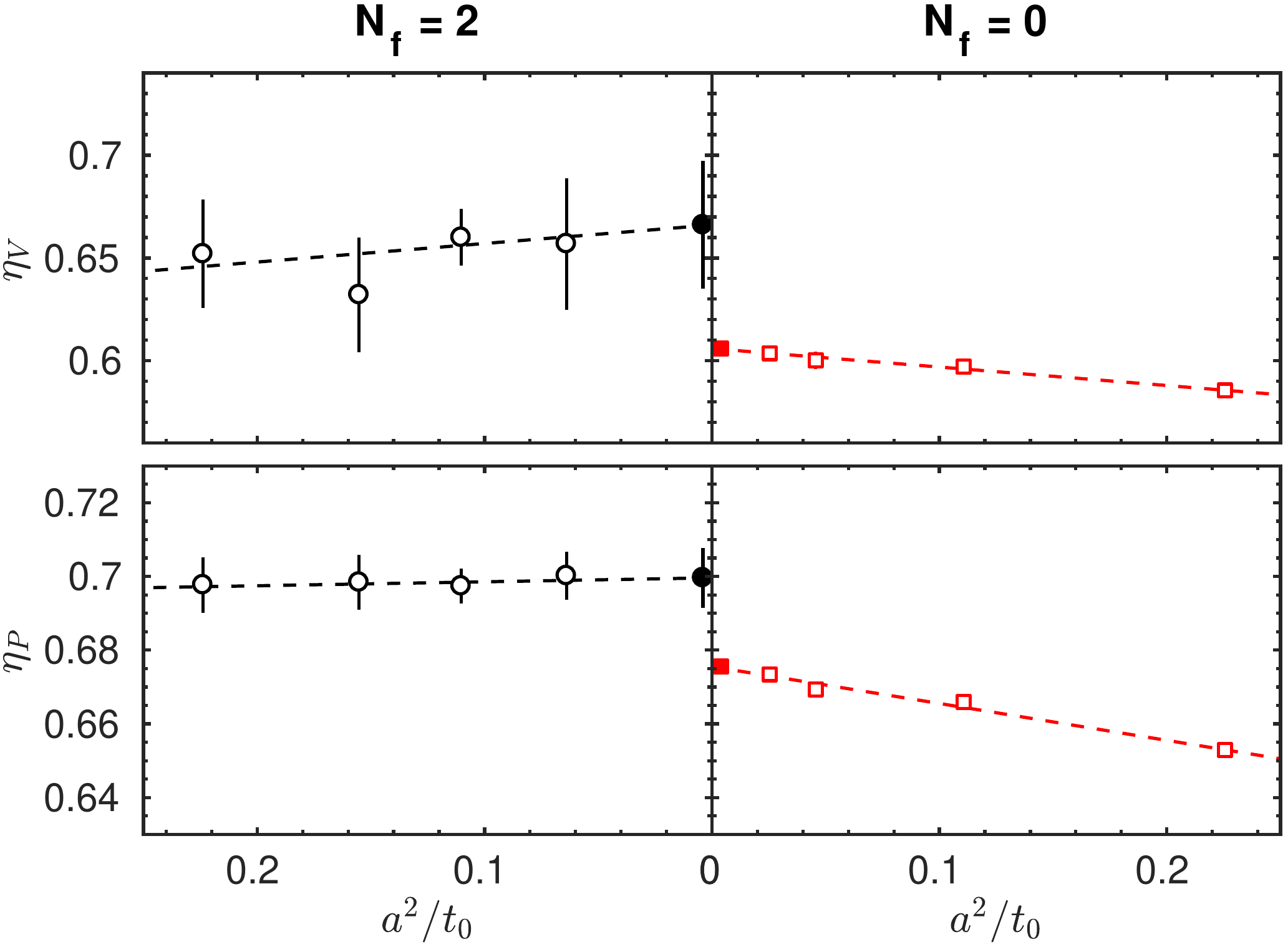}
   \caption{Continuum limits of the mass scaling functions $\eta_P$ and
   $\eta_V$ in both the $\Nf=2$ (left) and $N_f=0$ (right) theories.}\label{fig:eta}
\end{figure}

\begin{figure}[h]
   \centering
   \includegraphics[width=0.8\linewidth]{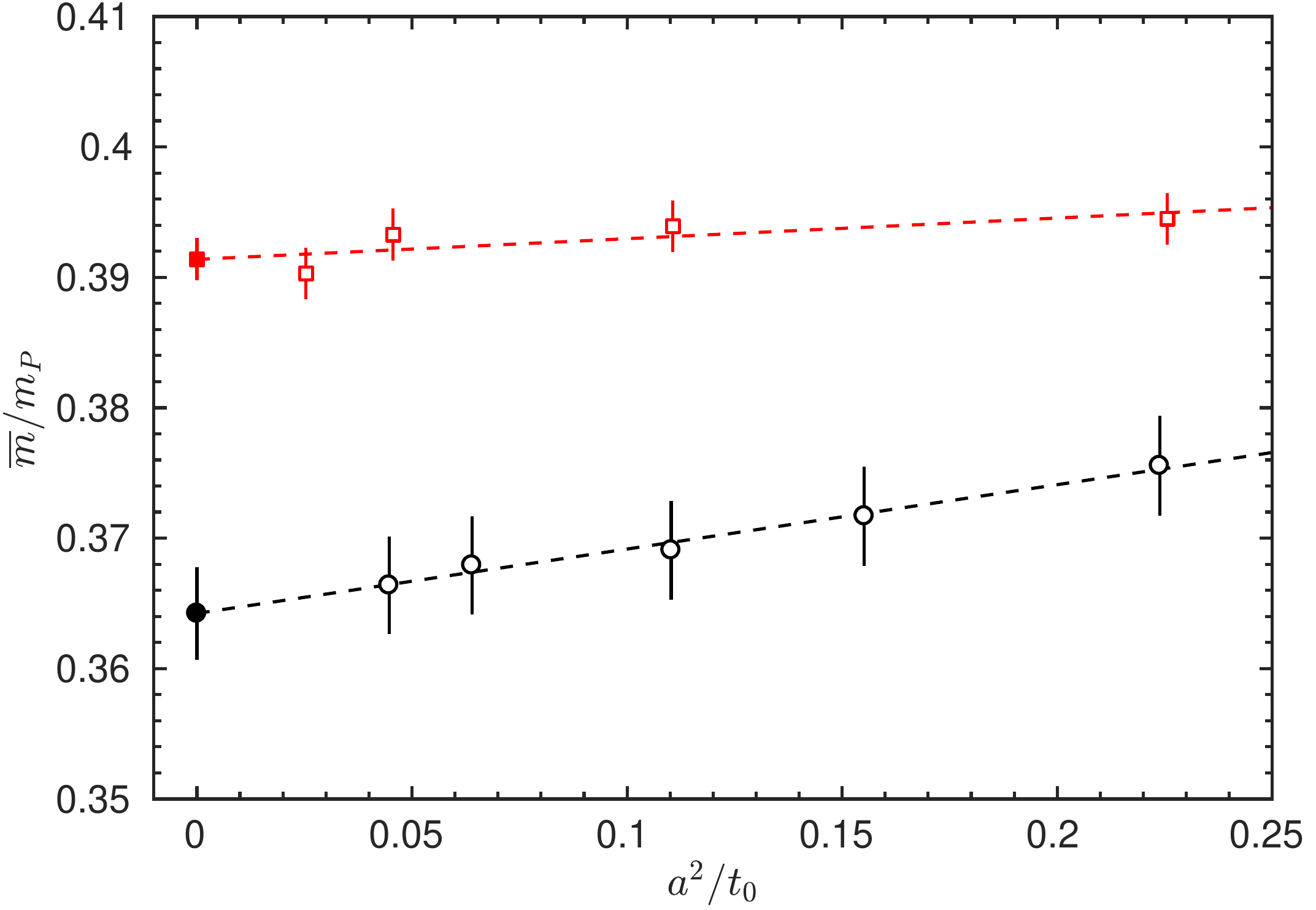}
   \caption{Continuum limits of the renormalized quark masses in the SF-scheme divided
   by the pseudoscalar meson mass. Note that the quark masses in 
            $\Nf=2$ and $\Nf=0$ theories are renormalized at different renormalization scales 
            and cannot be compared directly.}\label{fig:mbar}
\end{figure}

\Fig{fig:spectrum}-\Fig{fig:mbar} and \Tab{tab:contextrap} summarize our findings.
The data entering the fits are collected in \Tab{tab:ratios}.

The results in the continuum limit are collected in \Tab{tab:contextrap}
\begin{table}
\centering
\begin{tabular}{l l l l}
\toprule
Quantity        & $\Nf=2$     & $\Nf=0$     & sea effects [\%] \\
\midrule
$m_V/m_P$       & 1.05405(60) & 1.05274(46) & 0.124(71)\\
$m_S/m_P$       & 1.258(14)   & 1.224(20)   & 2.7(1.9)\\
$m_T/m_P$       & 1.271(38)   & 1.321(33)   & 3.9(4.1)\\
\midrule
$\eta_P$        & 0.6996(81)  & 0.67553(42) & 3.4(1.1)\\
$\eta_V$        & 0.666(31)   & 0.6060(13)  & 9.0(4.2)\\
\midrule
$M_c/m_P$       & 0.4764(74)  & 0.4528(51)  & 5.0(1.8)\\
\bottomrule
\end{tabular}
\caption{Results for various quantities in the continuum limit for both the 
$\Nf=0$ and the $\Nf=2$ theory.}\label{tab:contextrap}
\end{table}

\subsection{Dynamical charm effects}
The comparison of continuum results in the $\Nf=2$ theory with those in the $\Nf=0$
theory
directly quantifies the typical size of the effects, that the inclusion of dynamical 
charm quarks have on observables with valence charm quarks.

Although they were determined very precisely, no significant effect can be seen
in the meson mass spectrum. The most significant deviations of around $1.6\sigma$
are found in the ratios $m_V/m_P$ and $m_S/m_P$.
The relative differences between the central values of the first ratio are only 
$([m_V/m_P]^{\Nf=2} - [m_V/m_P]^{\Nf=0})/[m_V/m_P]^{\Nf=2} = 0.12(7)$\%.
For the hyperfine splitting $(m_V-m_P)/m_P$ this means a charm quark effect
of around 2\%.
In the $m_S/m_P$ ratio the central values deviate by $2.7(1.6)\%$.

A clearer difference between the $N_f=0$ and $N_f=2$ theories can be observed
in the mass-scaling functions and in the RGI quark mass. The values of $\eta_P$
and the quark mass differ by almost $3\sigma$. The relative differences are
$(\eta_P^{\Nf=2}-\eta_P^{\Nf=0})/\eta_P^{\Nf=2} = 3.4(1.1) \%$ and
$([M_c/m_P]^{\Nf=2}-[M_c/m_P]^{\Nf=0})/[M_c/m_P]^{\Nf=2} = 5.0(1.8)\%$. An even
larger (but less significant) difference is found in $\eta_V$.

\subsection{Lattice Artifacts}
\begin{figure}[h]
   \centering
   \includegraphics[width=0.8\linewidth]{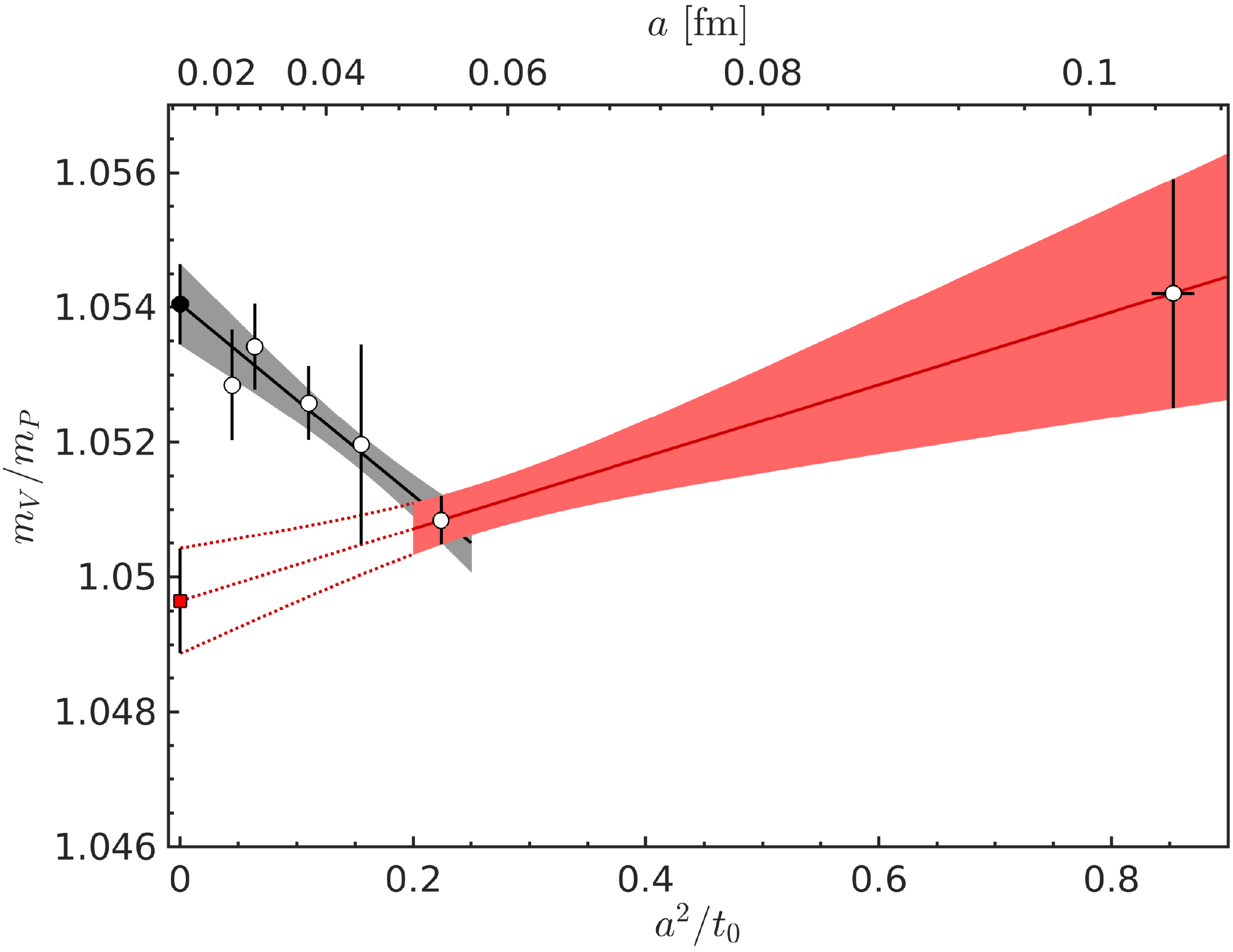}
   \caption{Continuum extrapolations of $m_V/m_P$. One extrapolation includes only data
   with $a^2/t_0 < 0.25$, the other uses only coarse lattices with $a^2/t_0>0.2$. 
   The continuum limits differ significantly between the two extrapolations.}\label{fig:latart}
\end{figure}

Having access to very fine lattice spacings is crucial for reliable continuum 
extrapolations. Although our fermionic action, i.e. twisted mass fermions 
with an additional clover term, is known to have relatively mild lattice 
artifacts, the continuum value of e.g. $m_V/m_P$ would be significantly
underestimated if we had access only to our two coarsest lattices (E and N). The finer
of the two has a lattice spacing of $a\approx 0.049$ fm, which is comparable 
to the finest lattice spacings typically achievable in large-volume simulations with 
light quarks. The situation is depicted in \Fig{fig:latart}.

The presence of large lattice artifacts of $O( (a\mu)^2 )$ not only affects 
observables like $m_V/m_P$, but also the value of the lattice spacing $a$ itself.
Since it is obtained by determining some hadronic length scale $L^{\rm had}/a$
in lattice units at finite lattice spacing and dividing it by the continuum value in fm, i.e.
$a = a/L^{\rm had} \times L^{\rm had, cont}$, its value depends on the 
lattice artifacts present in $L^{\rm had}$. In our case one possibility to 
compute the lattice spacings is through the 
scale $L^{\rm had, cont, 1} \equiv L_1 = 0.40(1)$ fm. Its values in lattice units
are known for our bare couplings and the resulting lattice spacings are between
$a^{L_1} = 0.023$ fm on ensemble $W$ and $a^{L_1} = 0.066$ fm on ensemble $E$.
Alternatively, one could determine the lattice spacing through
$L^{\rm had, cont, 2} \equiv \sqrt{t_0(M)} = 0.1131(38)$ fm~\cite{Athenodorou:2018wpk}. While
the two lattice spacing determinations agree well on the fine ensembles, the 
difference is quite substantial on the coarsest one, where we find 
$a^{t_0} \approx 0.1$ fm, i.e. we observe a 37\% lattice artifact in $a$!
Since $t_0/a^2$ is also determined on the quenched ensembles, we can determine
their lattice spacings using the decoupling relation 
$\sqrt{t_0(M)}^{\Nf=2} = \sqrt{t_0}^{\Nf=0} + O(M^{-2})$. Note that lattice spacings
determined by using the $N_f=0$ theory as an effective theory for our massive
two flavor theory differ from those determined by using it as an (uncontrolled) 
approximation to full QCD. In particular these lattice spacings depend on the 
value of $M$ in the fundamental theory. \Tab{tab:a} summarizes our scale setting.

\begin{table}[h!]
\centering
\begin{tabular}{c c c}
 \toprule
 Ensemble & $a^{L_1}$ [fm]    & $a^{t_0}$ [fm]\\
 \midrule
 E        & 0.066             & 0.104         \\
 N        & 0.049             & 0.054         \\
 O        & 0.042             & 0.045         \\
 P        & 0.036             & 0.038         \\
 S        & 0.028             & 0.029         \\
 W        & 0.023             & 0.024         \\
 \midrule
 qN       & -                 & 0.054         \\
 qP       & -                 & 0.038         \\
 qW       & -                 & 0.024         \\
 qX       & -                 & 0.018         \\
 \bottomrule
\end{tabular}
\caption{Lattice spacings in physical units on our quenched and dynamical ensembles,
determined in two different ways.}\label{tab:a}
\end{table}

\section{Conclusions}\label{s:conclusions}
In this work we presented a determination of the effects of charm quarks in
the sea based on a simulation of a model, QCD with $\nf=2$ charm quarks.
By comparing to the $\nf=0$ pure gauge theory at the matching point
defined in \eq{e:tuning} we can compute the size of these effects.
We find that they are below 2\% for the hyperfine splitting of charmonium.
These are good news for lattice QCD computations of charmonium based on
simulations of $\nf=2+1$ light quarks in the sea.
We also demonstrate in \fig{fig:latart} that lattice
spacings $a<0.06~$fm are needed for
safe continuum extrapolations of the charmonium spectrum
when using O($a$) improved Wilson quarks.

We also computed the effects of sea charm quarks in the mass-scaling
function $\eta$ of the charmonium masses \eq{e:etargi_ccbar}
and in the renormalization group invariant charm-quark mass $\Mc$.
\Tab{tab:contextrap} lists the comparison in the continuum limit
of these quantities in the $\nf=2$ and $\nf=0$ theory.
The effects of the charm sea quarks are clearly resolved
and their size is 3\% for $\eta_P$ and 5\% for $\Mc$. We notice that
our results are upper bounds for the effects of a charm sea quark
in QCD since in our model we have doubled their number.

Further analysis to compute charm loop effects in decay constants and
finestructure of $B_c$ mesons is in progress.
So far the disconnected contributions due to charm annihilation
\cite{Levkova:2010ft} have been
neglected since we computed isovector charmonium masses in our model.
Work on these contributions is under way.

\section*{Acknowledgments}

We thank our colleagues in the ALPHA collaboration for access to data analysis 
tools.
We gratefully acknowledge the computer resources
granted by the John von Neumann Institute for Computing (NIC)
and provided on the supercomputer JUROPA at J\"ulich
Supercomputing Centre (JSC) and by the Gauss Centre for
Supercomputing (GCS) through the NIC on the GCS share
of the supercomputer JUQUEEN at JSC,
with funding by the German Federal Ministry of Education and Research
(BMBF) and the German State Ministries for Research
of Baden-W\"urttemberg (MWK), Bayern (StMWFK) and
Nordrhein-Westfalen (MIWF).
S.C. acknowledges support from the European Union's Horizon 2020 research and innovation programme
under the Marie Sk\l{}odowska-Curie grant agreement No. 642069.

\appendix

\section{Parity and time-reflection symmetries}\label{s:parity}
Following transformations can be considered as a 
change of variables in the lattice path integral:

Parity
\begin{equation}
   \mathcal{P}: \begin{cases}
                   U_0(x_0,\vec x)      &\to U_0(x_0,-\vec x) \\
                   U_k(x_0,\vec x)      &\to U^\dagger_k(x_0,-\vec x -a \hat k),\qquad k=1,2,3 \\
                   \chi(x_0,\vec x)     &\to \gamma_0 \chi(x_0,-\vec x) \\
                   \bar\chi(x_0,\vec x) &\to \bar \chi(x_0, -\vec x) \gamma_0
                \end{cases}
\end{equation}

Time reflection
\begin{equation}
   \mathcal{T}: \begin{cases}
                   U_0(x_0,\vec x)      &\to U^\dagger_0(T-x_0-a,\vec x) \\
                   U_k(x_0,\vec x)      &\to U_k(T-x_0,\vec x),\qquad k=1,2,3 \\
                   \chi(x_0,\vec x)     &\to \gamma_0\gamma_5 \chi(T-x_0,\vec x) \\
                   \bar\chi(x_0,\vec x) &\to \bar \chi(T-x_0, \vec x) \gamma_5\gamma_0
                \end{cases}
\end{equation}

They are symmetries of the twisted mass action only if $\mu=0$. In general
these transformations lead to relations between expectation values in theories with positive
and with negative twisted masses. E.g. for the two-point functions like~\eq{eq:mesoncorr}
one finds
\begin{eqnarray}
   \langle \Op_1 \Op_2\rangle &=& \langle \mathcal{P}[\Op_1]\mathcal{P}[\Op_2]\rangle_{-\mu} \\
   \langle \Op_1 \Op_2\rangle &=& \langle \mathcal{T}[\Op_1]\mathcal{T}[\Op_2]\rangle_{-\mu}\, .
\end{eqnarray}
With standard Wilson fermions ($\mu=0$), these equations can be used to show that
$\langle \Op_1 \Op_2\rangle = 0$ if the operators $\Op_1$ and $\Op_2$ have opposite parity,
i.e. if $\mathcal{P}[\Op_1]\mathcal{P}[\Op_2] = -\Op_1 \Op_2$. As a consequence, 
an operator with a definite parity will only excite states with the same parity.
This property is lost in the twisted mass formulation, and in general operators
will excite states with both parities.

The combined $\mathcal{TP}$ transformation is a symmetry of the twisted mass 
action
\begin{equation}
   \langle \Op_1 \Op_2\rangle = \langle \mathcal{TP}[\Op_1]\mathcal{TP}[\Op_2]\rangle \\
\end{equation}
For the averaged correlators~\eq{eq:avcorr1}-\eq{eq:avcorr5} this means
\begin{equation}
   \frac{1}{2}\langle \Op_1\Op_2 + \mathcal{T}[\Op_1]\mathcal{T}[\Op_2]\rangle
   = \frac{1}{2}\langle \mathcal{TP}[\Op_1]\mathcal{TP}[\Op_2] + \mathcal{P}[\Op_1]\mathcal{P}[\Op_2]\rangle\, .
\end{equation}
It is now easy to see that the averaged correlator vanishes, if 
$\mathcal{P}[\Op_1]\mathcal{P}[\Op_2] = -\Op_1 \Op_2$. So, by enforcing the 
continuum time reflection symmetry of the correlator, automatically the mixing of
opposite parity operators is prohibited.

\newpage
\section{Tables}\label{sec:tables}
\begin{table}[h!]
\centering
\begin{tabular}{c l l l l l l}
\toprule
Ensemble & $a\mu$ & $am_P$ & $am_V$ & $am_T$ & $am_S$ & $t_0/a^2$ \\
\midrule
E        & 0.3809(54) & 1.667(16) & 1.757(18) & - & - & 1.172(25) \\
         & 0.36151 & 1.61079(13) & 1.69494(36) & - & - & 1.23907(82) \\
N        & 0.16647(28) & 0.8551(12) & 0.8985(12) & 1.005(51) & 0.977(15) & 4.468(12) \\
         & 0.166 & 0.85337(17) & 0.89687(34) & 1.000(51) & 0.975(15) & 4.4730(93) \\
O        & 0.13714(31) & 0.7117(13) & 0.7488(15) & - & - & 6.445(25) \\
         & 0.13095 & 0.68929(11) & 0.72742(26) & - & - & 6.561(12) \\
P        & 0.11482(32) & 0.6001(14) & 0.6317(15) & 0.707(19) & 0.7337(53) & 9.070(42) \\
         & 0.1132 & 0.59421(24) & 0.62579(40) & 0.703(20) & 0.7269(53) & 9.105(35) \\
S        & 0.08717(25) & 0.4570(10) & 0.4814(12) & - & - & 15.641(71) \\
         & 0.087626 & 0.45870(18) & 0.48309(34) & - & - & 15.621(60) \\
W        & 0.072557 & 0.38200(16) & 0.40219(31) & 0.481(10) & 0.4680(37) & 22.39(11) \\
\midrule
qN        & 0.17632(11) & 0.85846(37) & 0.90106(52) & 0.994(33) & 1.018(18) & 4.4329(38) \\
          & 0.16 & 0.806508(100) & 0.85216(48) & 0.933(38) & 0.972(22) & - \\
          & 0.17 & 0.838522(96) & 0.88224(41) & 0.970(35) & 1.000(20) & - \\
          & 0.18 & 0.870096(93) & 0.91204(36) & 1.008(32) & 1.028(18) & - \\
qP        & 0.12235(26) & 0.60125(100) & 0.6328(11) & 0.762(43) & 0.710(14) & 9.037(30) \\
          & 0.11 & 0.56074(32) & 0.59454(81) & 0.728(51) & 0.669(15) & - \\
          & 0.12 & 0.59373(31) & 0.62562(69) & 0.756(44) & 0.703(14) & - \\
          & 0.13 & 0.62619(30) & 0.65634(61) & 0.783(38) & 0.736(12) & - \\
qW        & 0.07798(19) & 0.38602(73) & 0.40607(100) & 0.4960(98) & 0.4713(63) & 21.925(83) \\
          & 0.07 & 0.35943(14) & 0.38103(77) & 0.472(11) & 0.4465(72) & - \\
          & 0.08 & 0.39294(14) & 0.41247(57) & 0.5022(94) & 0.4775(60) & - \\
          & 0.09 & 0.42573(13) & 0.44361(44) & 0.5322(79) & 0.5086(52) & - \\
qX        & 0.05771(13) & 0.28792(53) & 0.30296(56) & - & - & 39.41(14) \\
          & 0.056 & 0.28218(10) & 0.29755(20) & - & - & - \\
          & 0.058 & 0.288920(99) & 0.30389(19) & - & - & - \\
          & 0.06 & 0.295621(98) & 0.31022(19) & - & - & - \\
\bottomrule
\end{tabular}
\caption{Meson masses and $t_0$ in lattice units.
For $\Nf=2$ simulations the first line contains the values
extrapolated to the tuning point $\mu^\star$ and the second line
the values at the simulated parameters.
For $\Nf=0$ ensembles the first line contains the values interpolated
to $\mu^\star$, and the following three lines contain the values
	measured at different valence quark masses.}\label{tab:masses}
\end{table}

\begin{table}[h!]
\centering
\begin{tabular}{c l l l l l l}
\toprule
Ensemble & $m_V/m_P$ & $m_S/m_P$ & $m_T/m_P$ & $\overline{m}/m_P$ & $\eta_P$  & $\eta_V$ \\
\midrule
E        & 1.0542(17) & - & - & 0.4407(46) & 0.664(14) & 0.698(33) \\
N        & 1.05084(36) & 1.143(18) & 1.175(60) & 0.3756(38) & 0.6977(75) & 0.652(26) \\
O        & 1.0520(15) & - & - & 0.3717(38) & 0.6984(74) & 0.632(28) \\
P        & 1.05258(55) & 1.2226(86) & 1.177(31) & 0.3691(38) & 0.6974(47) & 0.660(14) \\
S        & 1.05342(64) & - & - & 0.3679(38) & 0.7002(65) & 0.657(32) \\
W        & 1.05285(82) & 1.225(12) & 1.259(29) & 0.3664(37) & - & - \\
\midrule
qN        & 1.04968(42) & 1.186(21) & 1.158(39) & 0.3945(20) & 0.65290(20) & 0.5856(16) \\
qP        & 1.0526(10) & 1.182(22) & 1.269(71) & 0.3939(20) & 0.66593(59) & 0.5972(29) \\
qW        & 1.0523(15) & 1.222(17) & 1.286(26) & 0.3933(20) & 0.66936(59) & 0.6003(42) \\
qX        & 1.05226(42) & - & - & 0.3903(20) & 0.67342(50) & 0.6037(12) \\
\bottomrule
\end{tabular}
\caption{Ratios of masses and the mass scaling functions.
All values are at the tuning point $\mu^\star$.}
\label{tab:ratios}
\end{table}

\newpage
\bibliographystyle{JHEP-2}
\bibliography{charm}
\end{document}